\documentclass[11pt]{revtex4-2}

\usepackage{hyperref}
\usepackage{braket}
\usepackage{amsmath,amsfonts,amssymb}
\hypersetup{dvips,dvipdfm,colorlinks=true,urlcolor=magenta,filecolor=magenta,linktoc=page,citecolor=red,linkcolor=blue,bookmarks=true}
\usepackage{graphicx,epstopdf}

\begin{document}
	\title{\textbf{Quantum Cosmology in Coupled Brans-Dicke Gravity: A Noether Symmetry Analysis}}
\author{ Dipankar Laya$^1$\footnote {dipankarlaya@gmail.com} Sourav Dutta$^2$\footnote {sduttaju@gmail.com} Subenoy Chakraborty$^1$\footnote {schakraborty.math@gmail.com}}
\affiliation{$^1$Department of Mathematics, Jadavpur University, Kolkata-700032, West Bengal, India\\$^2$Department of Mathematics, Dr. Meghnad Saha College, Itahar, Uttar Dinajpur-733128, West Bengal, India.}

\begin{abstract}
	The present work deals with a multi-field cosmological model in a spatially flat FLRW space-time geometry. The usual Brans-Dicke(BD) field and another scalar field are minimally coupled to gravity while they interact with  each	other through the Kinetic terms. {The main aim of the present work is to examine whether the model is compatible  with cosmic observations. So cosmological solutions are obtained using symmetry analysis only.} By imposing Noether Symmetry to the Lagrangian of the system the potential of the scalar field as well as the 
	coupling function has been determined. The classical solutions are 
	determined after simplifying the Lagrangian using cyclic variables. Finally, Wheeler-DeWitt(WD) equation in quantum cosmology has been formulated and conserved momenta corresponding to Noether symmetry shows the periodic part of the wave function and it helps to have the complete  integral for the wave function.
\end{abstract}

\maketitle


  \section{
  	Introduction}
  The Brans-Dicke (BD) scalar field theory is a modified theory of gravity, first introduced by Brans and Dicke \cite{r1,r2,r3} using Mach's principle\cite{r1} in the gravity 
  theory. In this picture the gravitational field is characterized not only geometrically but also by a non-minimally coupled scalar, known as Brans-Dicke(BD) scalar field. The coupling parameter (known as BD parameter) is considered as an effective Newton's constant and it measures the strength of the coupling between the scalar field and gravity. An extension of this scalar field model is the chiral model \cite{r4, r5, r6}  where two scalar fields are minimally coupled to gravity but the two scalar fields interact in the kinetic term. So one can 
  define a second-rank tensor from the Kinetic components of the scalar fields and for the chiral model it is a 2D hyperbolic sphere. Hence there does not exist any co-ordinate choice in which the two scalar fields are non-interacting. Such coupled two scalar 
  field models are widely used in the literature \cite{r7, r8, r9} and by 
  suitable extension it is possible to have the negative energy density of the scalar field \cite{r10, r11}. On the other hand, due to complexity of the modified Einstein field equations it become almost impossible to have an analytic solution.
   So normally, one has to use qualitative tools in the theory of differential equations to analyzed these physical theories. In particular, the group invariant transformations method is suitable in the present problem. More precisely, geometric symmetries related to space-time i.e., Noether symmetry is very 
  much relevant to analyze the BD cosmological model. The advantage of using  Noether symmetry is of two fold : the associated conserved quantities  (charge)can be used as a selection criterion to determine similar physical processes and the symmetry analysis either simplify the evolution equations or determine the integrability of  the system. Further, Noether symmetry analysis examines the self consistency of the phenomenological 
  physical models and the physical parameters (involved in the physical system) are contained by this symmetry analysis \cite{r23, r24, r25,r44,r45, r26, r27, r28, r29}.
    The symmetry vector identifies a transformation in the augmented space so that one of the variables becomes cyclic and consequently the Lagrangian simplifies to a great extend \cite{r36, r37} and the evolution 
  equation becomes solvable. In quantum cosmology the Wheeler-DeWitt(WD) equation has been formed and Noether charges identify the oscillatory part 
  of the solution.Then the WD equation becomes easily solvable.\\
  
  {In cosmology, scalar field models may describe the early inflationary era  (as inflaton field) or may be responsible for the late time accelerating phase (as a DE candidate). The motivation of the present two scalar fields model is to examine whether the model is in accord with the cosmological observations. Due to complicated nature of the model it is not possible to have analytic classical solution by solving the usual field equations. Noether symmetry analysis provides the tool  for solving the field equations. For quantum cosmology, the motivation is to examine whether the initial singularity may be avoided by quantum description. The present work deals with both classical and quantum 
  cosmological study of two scalar fields coupled BD theory of gravity by Noether symmetry analysis.} The plan of the 
  work is as follows: Section II deals with an overview of the Noether symmetry analysis. Section III presents the mathematical description of the coupled two scalar 
  field BD theory and Noether symmetry has been imposed to have the 
  symmetric vector. Also the coupled Lagrangian and the evolution equations are simplified by identifying the cyclic variable in the augmented space. Quantum cosmology of the present model has been formulated in section IV. The conserved Noether charge identifies the oscillatory part of the wave function and as a result the WD equation has been solved in this section. The paper ends with a brief discussion and concluding remarks.   
  \section{
  	Noether Symmetry Analysis : An overview}
  According to Noether's 1st \cite{r36, r37, r35} 
  theorem every physical 
  system is associated to some conserved quantities provided the Lagrangian of the system is invariant with respect to the Lie derivative along the appropriate vector field i.e., Mathematically 
  \begin{equation}
  \mathcal{L_{\overrightarrow{X}}}L=\overrightarrow{X}(L)=0
  \end{equation}
  Thus for a point like canonical Lagrangian 
  $L=L[q^{\alpha}(x^i),\partial_jq^{\alpha}(x^i)],~q^{\alpha}(x^i)$ being the 
  generalized  co-ordinate, the Euler-Lagrange equations are
  \begin{equation}
  \partial_j\left(\frac{\partial 
  	L}{\partial(\partial_jq^{\alpha})}\right)-\frac{\partial L}{\partial 
  	q^{\alpha}}=0
  \end{equation}
  Now contracting with some unknown function $\lambda^{\alpha}(q^{\beta})$ i.e.,
  \begin{equation}
  \lambda^{\alpha}\Big\{	\partial_j\Big(\frac{\partial 
  	L}{\partial(\partial_jq^{\alpha})}\Big)-\frac{\partial L}{\partial 
  	q^{\alpha}}\Big\}=0\nonumber
  \end{equation}
   gives after simplification
  \begin{equation}
  \lambda^{\alpha}\frac{\partial L}{\partial 
  	q^{\alpha}} + (\partial_j\lambda^{\alpha})	
  \frac{\partial 
  	L}{\partial(\partial_jq^{\alpha})} = 
  \partial_j\Big(\lambda^{\alpha}	
  \frac{\partial 
  	L}{\partial(\partial_jq^{\alpha})}\Big)\nonumber
  \end{equation}
  Thus 
  \begin{equation}
  \mathcal{L_{\overrightarrow{X}}}L= \lambda^{\alpha}\frac{\partial L}{\partial 
  	q^{\alpha}} + (\partial_j\lambda^{\alpha})	
  \frac{\partial 
  	L}{\partial(\partial_jq^{\alpha})} = 
  \partial_j\Big(\lambda^{\alpha}	
  \frac{\partial 
  	L}{\partial(\partial_jq^{\alpha})}\Big)
  \end{equation}
  will provided \cite{r16, r17}
  \begin{equation}
  Q^j=\lambda^{\alpha}	\frac{\partial L}{\partial(\partial_jq^{\alpha})}
  \end{equation} 
  to be conserved (i.e., $\partial_jQ^j=0$) and 
  \begin{equation}
  \overrightarrow{X}= \lambda^{\alpha}\frac{\partial}{\partial 
  	q^{\alpha}} + (\partial_j\lambda^{\alpha})	
  \frac{\partial }{\partial(\partial_jq^{\alpha})}
  \end{equation}
  is the symmetry vector.
  Now the energy function associated with the system can be written as 
  \begin{equation}
  E=\dot{q}^{\alpha}\frac{\partial L}{\partial\dot{q}^{\alpha}}-L
  \end{equation}
  Thus associated with Noether symmetry there is a conserved 
  current $Q^j$ (given 
  in equation (4)). The scalar quantity obtained by integrating the time 
  component of $Q^j$ over the spatial  volume is termed as (conserved) Noether charge $(Q)$. As in the present homogeneous geometry all the variables are 
  only time dependent so the Noether current defined in equation $(4)$ coincides with 
  the Noether charge. Further, the Noether charge can be geometrically 
  interpreted as the inner product of the infinitesimal generator with cartan one 
  form \cite{r44a} as 
  \begin{equation}
  Q=i_{\overrightarrow{X}}\theta_L
  \end{equation} 
  where $i_{\overrightarrow{X}}$ indicates the inner product with vector field 
  $\overrightarrow{X}$ and 
  \begin{equation}
  \theta_L=\frac{\partial L}{\partial q^{\alpha}}dq^{\alpha}
  \end{equation}
  is the cartan one form.
  
  Further, this geometric inner product representation is suitable to 
  identify cyclic variables in the augmented space . If 
  $q^{\alpha}\rightarrow s^{\alpha}$ is a transformation in the augmented 
  space then the symmetry vector transforms to 
  \begin{equation}
  \overrightarrow{X}_T=(i_{\overrightarrow{X}}ds^{\alpha})
  \frac{\partial}{\partial s^{\alpha}}+
  \Big\{{\frac{d}{dt}(i_{\overrightarrow{X}}ds^{s})}\Big\}\frac{d}{ds^{\alpha}}
  \end{equation}
  This transformed symmetry vector $\overrightarrow{X}_T$ is nothing 
  but a lift of the vector field $~X~$ on the augmented space. Now, without loss 
  of generality if the above point like transformation is restricted to 
  \begin{equation}
  i_{\overrightarrow{X}_T}ds^{\alpha}=1,~\mbox{for}~\alpha=m ~~ 
  \mbox{(say)}\nonumber
  \end{equation}
  \begin{equation}
  i_{\overrightarrow{X}_T}ds^{\alpha}=0,~\mbox{for}~ 
  \alpha\neq m
  \end{equation}
  then
  \begin{equation}
  \overrightarrow{X}_T=\frac{\partial}{\partial s^m} ~~\mbox{and}~~
  \frac{\partial L_T}{\partial s^m}=0
  \end{equation}
  Thus augmented variable $s^m$ is a cyclic variable . The above geometric 
  process can be interpreted as to choose the transformed infinitesimal generator 
  along the co-ordinate line (identified as the cyclic variable).
  
  Moreover, 
  if the Lagrangian of the system has no explicit time dependence then the above 
  energy function (given in equation (6)) is nothing but the Hamiltonian of the 
  system and it is also a constant of motion \cite{r37} . The Hamiltonian formulation is 
  very suitable in the context of quantum cosmology. Then the Noether symmetry 
  condition modifies to 
  \begin{equation}
  \mathcal{L}_{\overrightarrow{X}_H}H=0
  \end{equation}
  With $~~\overrightarrow{X}_H=\dot{q}^{\alpha}\frac{\partial}{\partial 
  	q^{\alpha}}+\ddot{q}^{\alpha}\frac{\partial}{\partial \dot{q}^{\alpha}}$, as the symmetry vector. The conjugate  momenta which are conserved due to 
  Noether symmetry can be written as 
  \begin{equation}
  \pi_{\beta}=\frac{\partial L}{\partial q^{\beta}} =\Sigma_{\beta} 
 ,~~~\beta=1,2,...,r
  \end{equation}
  where $r$ is the number of symmetries. In quantization scheme the operator 
  version of the above conserved momentum takes the form 
  \begin{equation}
  -i\partial_{q^\beta}|\Psi>=\Sigma_{\beta}|\Psi>
  \end{equation}
  So for real $\Sigma_{\beta}$, the differential equation $(14)$ has the 
  solution 
  \begin{equation}
  |\Psi>=\sum_{\beta=1}^{r}e^{i\Sigma_{\beta}q^{\beta}}|\phi(q^k)>,~~k<n
  \end{equation}
  where `$k$' is the direction along which there is no symmetry and $n$ is the 
  dimension of the minisuperspace. Thus according to Hartle 
  \cite{rh}  the oscillatory
  part of the wave function implies the existence of the Noether symmetry and the 
  conjugate momentum along the symmetry direction is conserved. Therefore, 
  using Noether symmetry it is possible to consider entire class of hypothetical 
  Lagrangians with given invariant for description of a physical system .

  \section{\textbf{Multifield Cosmological Model and Noether Symmetry}}
  In multifield cosmological model, we consider two scalar fields which are 
  minimally coupled to gravity but the two scalar fields interact in the 
  kinetic term. So the action integral for this model is given by 		
  \begin{equation}
  \mathcal{A}=\int d^4x\sqrt{-g}\Big[\frac{\phi R}{2} + 
  \frac{\omega}{2\phi}g^{\mu\nu}\phi_{;\mu}\phi_{;\nu} + \frac{1}{2} 
  F^2(\phi)g^{\mu\nu}\psi_{;\mu}\psi_{;\nu} + V(\phi) \Big]
  \end{equation}
  where $~\phi(x^{\alpha})~$ is the usual BD scalar field, $~\omega~$ is the BD 
  coupling parameter, $~V(\phi)~$ is the potential function and 
  $~\psi(x^{\alpha})~$
  is the other scalar field minimally coupled to gravity and $~F(\phi)~$ is the 
  coupling function indicating interaction between the two scalar fields. In the 
  background of homogeneous and isotropic flat FLRW space-time model the 
  Lagrangian of the above cosmological model is given by \cite{rA}
  \begin{equation}
  L(a,\dot{a},\phi,\dot{\phi},\psi,\dot{\psi} )= 
  3a\dot{a}^2\phi+3a^2\dot{a}\dot{\phi}- \frac{\omega}{2\phi}a^3\dot\phi^2-
  \frac{a^3}{2}F^2(\phi)\dot\psi^2+a^3V(\phi)	
  \end{equation}
  So the field equations for the present cosmological model (obtained by Euler-
  Lagrange equations) are given by 
  \begin{equation}
  3H^2\Big(1+\frac{\dot{\phi}}{H\phi}\Big) = \frac{\omega}{2} 
  \Big(\frac{\dot{\phi}}{\phi}\Big)^2 + \frac{F^2(\phi)}{2\phi}\dot{\psi}^2 - 
  \frac{V(\phi)}{\phi}
  \end{equation}
  \begin{equation}
  2\dot{H} = -2H\frac{\dot{\phi}}{\phi} - 3H^2 - \frac{\omega}{2} 
  \Big(\frac{\dot{\phi}}{\phi}\Big)^2 - \frac{F^2(\phi)}{2\phi}\dot{\psi}^2 - 
  \frac{\ddot{\phi}}{\phi} + \frac{V(\phi)}{\phi}
  \end{equation}
  \begin{equation}
  \ddot{\phi}+3H\dot{\phi}-\frac{1}{2}\Big(\frac{\dot{\phi}}{\phi}\Big)^2 + 
  \frac{1}{\omega}\Big[ 6H^2\phi+3\dot{H}\phi + \phi V_{,\phi} - 
  \frac{1}{2}(F^2)_{,\phi}\phi\dot{\psi}^2 \Big]=0
  \end{equation}
  and
  \begin{equation}
  \ddot{\psi}+3H\dot{\psi}+(\ln(F^2))_{,\phi}\dot{\phi}\dot{\psi}=0
  \end{equation}
  Note that equation $(18)$ is known as the Hamiltonian constraint for the 
  present model and $~\psi~$ is a cyclic co-ordinate for the above Lagrangian .
  
  The infinitesimal generator corresponding to Noether symmetry 
  is given by 
  \begin{equation}
  \overrightarrow{X}=\alpha \frac{\partial}{\partial{a}} 
  +\beta \frac{\partial}{\partial{\phi}}+ 
  \dot{\alpha}\frac{\partial}{\partial{\dot{a}}}+ 
  \dot{\beta}\frac{\partial}{\partial\dot{\phi}}+ 
  \dot{\gamma}\frac{\partial}{\partial\dot{\psi}}
  \end{equation}
  with  $\alpha=\alpha(a,\phi,\psi)$ ,  $\beta=\beta(a,\phi,\psi)$ ,  
  $\gamma=\gamma(\psi)$  ,
  $\dot{\alpha}=\frac{\partial\alpha}{\partial 
  	a}\dot{a}+\frac{\partial\alpha}{\partial \phi}\dot{\phi}+
  \frac{\partial\alpha}{\partial\psi}\dot{\psi}$ , and similarly for 
  $\dot{\beta}$ and $\dot{\gamma}$.
  As due to Noether symmetry $ \mathcal{L}_{\overrightarrow{X}}L =0 
  \implies\overrightarrow{X} L=0$, so $\alpha, 
  ~\beta,$ and $\gamma$ will satisfy the 
  following set of equations 
  \begin{eqnarray}
  \phi\alpha+a\beta+2a\phi\frac{\partial\alpha}{\partial a} + 
  a^2\frac{\partial\beta}{\partial a} &=&0\\
  3a^2\phi\omega\alpha-a^3\omega\beta 
  -6a^2\phi^2\frac{\partial\alpha}{\partial\phi}
  + 2a^3\omega\phi\frac{\partial\beta}{\partial\phi}&=&0\\
  6a\phi\alpha+3a^2\phi\frac{\partial\alpha}{\partial 
  	a} + 6a\phi^2\frac{\partial\alpha}{\partial\phi} 
  +3a^2\phi\frac{\partial\beta}{\partial\phi}- 
  a^3\omega\frac{\partial\beta}{\partial a}&=&0\\
  3F(\phi)\alpha+2a\beta 
  F'(\phi)+ 2aF(\phi)\frac{\partial\gamma}{\partial\psi}&=&0\\
  2\phi\frac{\partial\alpha}{\partial\psi}+ 
  a\frac{\partial\beta}{\partial\psi}&=&0\\
  3\phi\frac{\partial\alpha}{\partial\psi} 
  -a\omega\frac{\partial\beta}{\partial\psi}&=&0\\
  3\alpha V(\phi)+a\beta V'(\phi)&=&0
  \end{eqnarray}
  The above set of seven equations(23--29) are first order 
  partial differential equations, to determine the infinitesimal generator (22) of the above Noether symmetry 
  and also the unknown coupling function $F(\phi)$ and potential function 
  $V(\phi)$, we use the method of separation of variables and the solution takes 
  the form 
  \begin{equation}
  \alpha=\alpha_0a,~\beta=\beta_0\phi,~\gamma=\gamma_0,~V=V_0\phi,~ F=F_0 
  \phi^{\frac{1}{2}} 
  \end{equation}
  where $~\alpha_0,~\beta_0$ are related with the relation $3\alpha_0+\beta_0=0,~\gamma_0,~V_0,~F_0~$ are integration constants. 
  
  Now, one makes a transformation $(a,\phi,\psi)\rightarrow(P,M,R)$ in the 
  augmented space so that one of the variables becomes cyclic. Using the 
  technique of inner product with cartan one form (defined in the previous section:  equation (10)) one gets  
  \begin{eqnarray}
  \ln\left(\frac{a}{\phi}\right)&=&(\alpha_0-\beta_0)P\nonumber\\
  \ln(a^3\phi)&=&M\nonumber\\
  \psi&=&\gamma_0R	
  \end{eqnarray}
  As a result, the Lagrangian of the system simplifies to
  \begin{equation}
  L=e^M\Big\{A\dot{P}^2+B\dot{M}^2+C\dot{P}\dot{M}+V_0-D\dot{R}^2\Big\}
  \end{equation}
  
  where the constants $~A,B,C,D~$ are connected with 
  $~\omega~$ by the following relations 
  \begin{eqnarray}
  A&=&(\alpha_0-\beta_0)^2\Big(-\frac{6}{16}-\frac{9\omega}{32}\Big) 
  \nonumber\\
  B&=&\Big(\frac{6}{16}-\frac{\omega}{32}\Big)\nonumber\\
  C&=&(\alpha_0-\beta_0)\frac{6\omega}{32}\nonumber\\
  D&=&\frac{F_0^2\gamma_0^2}{2}\nonumber
  \end{eqnarray}
  
  Now solving the evolution equations corresponding to this transformation (32) one can get:
  \begin{eqnarray}
  	P&=&\frac{k_1}{bA\Delta \sqrt{\Gamma}}F\left[\frac{\pi-2b(t+c_2)}{4}{\bigg|}\frac{2V_0 \Delta}{V_0 \Delta \pm 1}\right]\mp C_1 \ln\{1-\Delta \sin\{b(t+c_2)\}\}+c_4 \\
  M&=&\frac{1}{2} \ln\Big\{\Gamma(1\mp\Delta \sin b(t+c_2)\Big\}\\	
  R&=&\frac{-2k_2}{b\sqrt{\Gamma (V_0 \Delta \mp 1)}}F\left[\frac{\pi -2b(t+c_2)}{4}{\bigg|}\frac{2V_0 \Delta}{V_0 \Delta \pm 1}\right] +c_3 
  \end{eqnarray}
  
  where $ C_1=\dfrac{c}{4Ab \Delta \Gamma} \sqrt{4c_1 - 2V_0 \Gamma ^2 b^2 },~k_1,~\Delta,~\Gamma,~b,~c_1,~c_2,~c_3,~c_4,~k_2$ are 
  constants and  $F[x|y]$ is the elliptic integral of the first kind with 
  parameter $y=n^2~$(where $~n~$ being any real number) or in terms of the old 
  variables the solution becomes 
  {\scriptsize\begin{eqnarray}
  	a&=&\exp\left[\frac{1}{8}\ln\left(\Gamma \mp\Gamma\Delta\sin[b(t+c_2)]\right)+(\alpha_0 -\beta_0)\left\{\frac{k_1}{b\Delta A \sqrt{\Gamma }}F\left[\frac{\pi-2b(t+c_2)}{4}{\bigg|}\frac{2V_0 \Delta}{V_0 \Delta \mp 1}\right]\right.\right. \nonumber\\&&~~~~~\mp C_1\ln(1-\Delta \sin(b(t+c_2))) +c_4\bigg\}\bigg]\label{26}
  	\end{eqnarray}}
  {\scriptsize\begin{eqnarray}
  	\phi&=&\exp\left[\frac{1}{8}\ln\left(\Gamma \mp\Gamma\Delta\sin[b(t+c_2)]\right)-3(\alpha_0 -\beta_0)\left\{\frac{k_1}{b\Delta A \sqrt{\Gamma }}F\left[\frac{\pi-2b(t+c_2)}{4}{\bigg|}\frac{2V_0 \Delta}{V_0 \Delta \mp 1}\right]\right.\right. \nonumber\\&&~~~~~\mp C_1\ln(1-\Delta \sin(b(t+c_2))) +c_4\bigg\}\bigg]\label{27}
  	\end{eqnarray}}
  \begin{equation}
  \psi=-\frac{2k\gamma_0}{b\Gamma\sqrt{V_0 \Delta \mp 1}}F\Big[\frac{\pi-2b(t+c_2)}{4}\Big{|}\frac{2\Delta V_0}{\Delta V_0 \mp 1}\Big]+c_3
  \end{equation}
  
\begin{figure}[h]
  \begin{minipage}{0.47\textwidth}
  	\centering \includegraphics[height=5cm,width=8cm]{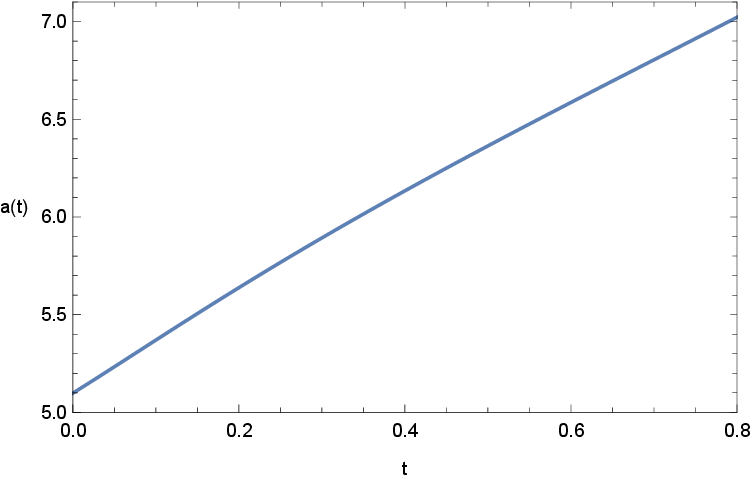}
  	\end{minipage}\hfill
  	 \begin{minipage}{0.47\textwidth}
  	\centering \includegraphics[height=5cm,width=8cm]{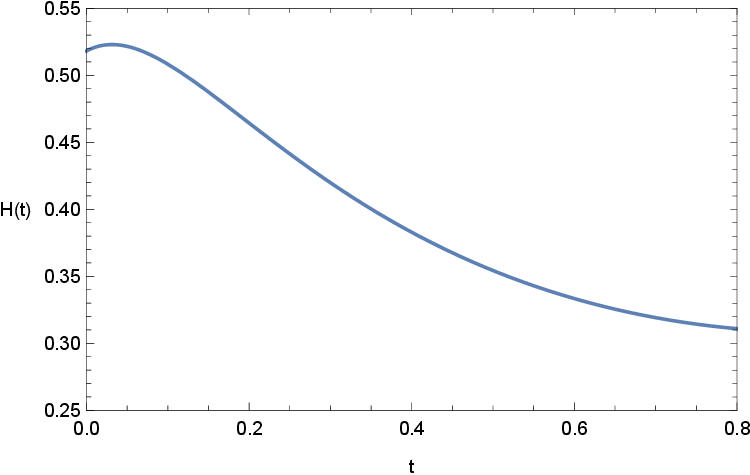}
  	\end{minipage}
  	\begin{minipage}{0.47\textwidth}
  	\centering \includegraphics[height=5cm,width=8cm]{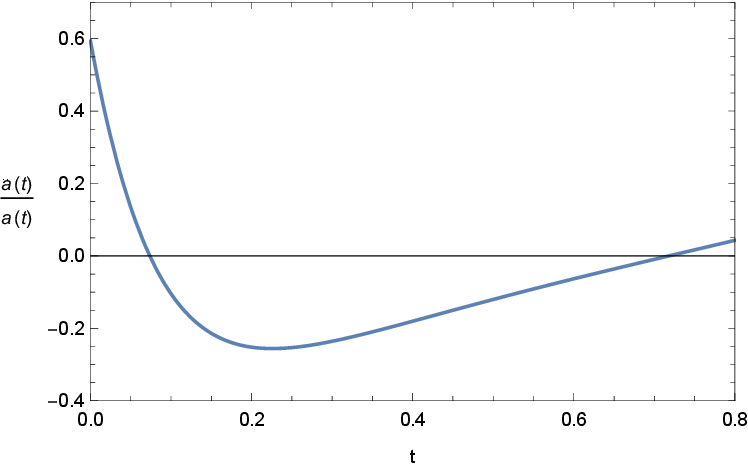}
  	\end{minipage}
  	\caption{The graphical representation of scale factor $a(t)$ (top left), Hubble parameter $H(t)$ (top right) and acceleration parameter $\frac{\ddot{a}(t)}{a(t)}$ (bottom) with respect to cosmic time $t$.}\label{f1}
  \end{figure}

  The solutions for the scale factor $~a~$ and the cosmological parameters namely 
  $H= \frac{\dot{a}}{a}$ and the acceleration parameter $~\frac{\ddot{a}}{a}~$ 
  have been plotted in FIG. \ref{f1} for various choices for the parameters 
  involved. Form the figures we see that the present cosmological model is an 
  expanding model of the universe and the Hubble parameter is positive but 
  decreases with time. The graph of the acceleration parameter show that the 
  universe was in an accelerated era of expansion at the early phase then there 
  was an epoch of decelerated expansion and finally the universe accelerated 
  again. So the cosmic evolution matches with the qualitative evolution as 
  predicted by observations. 
  
  
  
  On the other hand, the Lagrangian for the present model defines the kinetic 
  metric as
  \begin{equation}
  dS^2 _{(k)}=-6a\phi da^2 -6a^2dad\phi+\frac{\omega}{\phi}a^3 d\phi ^2 
  +a^3\{F(\phi)\}^2 d\psi ^2 
  \end{equation}
  with effective potential $a^3 V(\phi)$. Using the change of variables as
  \begin{equation}
  e^u=a~~,~~ e^v=\phi ~~i.e.,~~\frac{\dot{a}}{a}=\dot{u}~~,~~ 
  \frac{\dot{\phi}}{\phi}=\dot{v},\nonumber
  \end{equation}
  the above kinetic metric becomes
  \begin{eqnarray}
  dS^2 _{(k)}&=&a^3\phi\Big[-6du^2 -6dudv+\omega dv^2+F^2 _0 
  d\psi^2\Big]\nonumber\\ 
  &=&e^{(3u+v)}\Big[-6d\big(u+\frac{v}{2}\big)^2+\big(\omega
  -\frac{3}{2}\big)dv^2+F^2 _0 d\psi ^2\Big]\nonumber
  \end{eqnarray}
  Substituting $~\xi=12(u+\frac{v}{2}),~~2\eta=(\omega-\frac{3}{2})v,~~2\delta 
  =F_0 \psi~$ one gets
  
  \begin{equation}
  dS^2 _{(k)}=e^{(3u+v)} \big[-d\xi^2+d\eta^2+d\delta^2 \big]
  \end{equation}
  The kinetic metric is conformal to flat 3D metric. This 3D metric admits a 
  gradient homothetic vector field $~H_V =\frac{2}{3}a\partial _a ~,~ 
  \psi_{H_V}=1$  .
  
  Note that this homothetic vector field does not generate a Noether point 
  symmetry for the Lagrangian. However, additional Noether point symmetries are 
  possible for the present flat case due to three killing vectors of the 2D 
  metric and these killing vector fields span the $E(2)$ group.
  Further, the transformed Lagrangian takes the form 
  \begin{equation}
  L=-\frac{1}{2}\dot{\xi}^2 +\frac{1}{2}\dot{\eta}^2 + 
  \frac{1}{2}\dot{\delta}^2 + V_0, 
  \end{equation}
  which is nothing but the Lagrangian for the quintom dark energy model, Further 
  written $\Phi=\eta +i\delta$, the action can be equivalent to a single complex 
  scalar field.

  ~~~~~~~~~~~~~~~~~~~~~~~~~~~~~~~~~~~~~~~~~~~~~~~~~~~~~~~~~~~~~~~~~~~~~~~~~~~~~~~~
  \section{Quantum Cosmology and Noether Symmetry Analysis}
  In this section Noether symmetry analysis has been used to formulate quantum 
  cosmology of the present cosmological model. In particular, the conserved 
  quantities (namely Noether charge) corresponding to this symmetry determine the 
  oscillatory part of the wave function of the universe and as a result the 
  WD equation simplifies to a great extend and may be solvable.
  
  Usually in cosmology, one uses the simplest and widely used minisuperspace 
  \cite{Bhaumik:2022adi} 
  models corresponding to homogeneous and isotropic space-time metrics. So the 
  Lapse function $N(=N(t))$ is homogeneous and the shift function vanishes 
  identically. As a result, the 4D manifold has the metric in 
  (3+1)-decomposition as 
  \begin{equation}
  ds^2=-N(t)dt^2+q_{ab}(x,t)dx^adx^b~~,~~a,b=1,2,3.
  \end{equation} 
  In this (3+1) decomposition the Einstein-Hilbert action can be written as
  \begin{equation}
  I(q_{ab},N)=\int dtd^3xN\sqrt{q}[K_{ab}K^{ab}-K^2+{}^{(3)}R]
  \end{equation}
  with $K_{ab}$, the extrinsic curvature of the three space and $3_{(R)}$, the 
  curvature scalar of the three space.
  
  Now as the three space is homogeneous, so 
  the three metric $q_{ab}$ can be described by a finite number of time functions, 
  namely $q^{\alpha}(t),~~\alpha=0,1,2,...,n-1$. As a consequence, the above 
  action can be written as that of a relativistic point particle having self 
  interacting potential in $nD$ curved space-time as \cite{r42, r43} 
  \begin{equation}
  I(q^{\alpha},N)=\int_{0}^{1}Ndt\Big[\frac{1}{2N^2}\mu_{\alpha\beta}(q) 
  \dot{q}^{\alpha}\dot{q}^{\beta}-U(q) \Big],
  \end{equation} 
  with equation of motion of the relativistic particle as 
  \begin{equation}
  \frac{1}{N}\frac{d}{dt}\Big(\frac{\dot{q}^{\alpha}}{N}\Big) + \frac{1}{N^2} 
  \Gamma_{\beta\gamma}^{\alpha}\dot{q}^{\beta}\dot{q}^{\gamma} + 
  \mu^{\alpha\beta}\frac{\partial U}{\partial q^{\beta}}=0 .
  \end{equation}
  Also the variation of the action with respect to the Lapse function gives the 
  constraint equation (Hamiltonian constraints)
  \begin{equation}
  \frac{1}{2N^2}\mu_{\alpha\beta}\dot{q}^{\alpha}\dot{q}^{\beta}+V(q)=0
  \end{equation}
  Using the momenta conjugate to $q^{\alpha} ~~i.e.,~~p_{\alpha} = \frac{\partial 
  	L}{\partial\dot{q}}_{\alpha}=\mu_{\alpha\beta}\frac{\dot{q}^{\beta}}{N}$ the 
  Hamiltonian of the system turns out to be 
  \begin{equation}
  H=p_{\alpha}\dot{q}^{\alpha}-L=N\big[\frac{1}{2}\mu^{\alpha\beta}p_{\alpha} 
  p_{\beta}+V(q)\big]\equiv NH
  \end{equation}
  Now combining equations (46) and (47) one gets
  \begin{equation}
  H(q^{\alpha},p_{\alpha})\equiv\frac{1}{2}\mu^{\alpha\beta}p_{\alpha} 
  p_{\beta}+V(q)=0
  \end{equation}
  In quantization scheme, if one writes the operator version of the the above 
  Hamiltonian constraint then it becomes the Wheeler-DeWitt(WD) equation in 
  quantum cosmology as the form 
  \begin{equation}
  H\Big(q^{\alpha}-ih\frac{\partial}{\partial q^{\alpha}}\Big)\Psi(q^{\alpha})=0 	
  \end{equation}
  This is a second order hyperbolic type partial differential equation. The 
  problem related to formulation of these equations is the operator ordering 
  problem as the minisuperspace metric in general depends on $q^{\alpha}$. A 
  possible resolution of this problem is to impose the covariant nature of the 
  minisuperspace quantization.
  
  Further, in the quantization scheme of minisuperspace there 
  exists a conserved current for probability measure namely
  \begin{equation}
  \overrightarrow{J}=\frac{i}{2}(\Psi^{\star}\bigtriangledown\Psi - 
  \Psi\bigtriangledown\Psi^{\star})
  \end{equation} 
  with $ \overrightarrow{\bigtriangledown}.\overrightarrow{J}=0$. Here $~\Psi~$ 
  satisfies the above WD equation and probability measure can be express as 
  \begin{equation}
  dp=|\Psi(q^{\alpha})|^2dV
  \end{equation} 
  with $dV$ representing a volume element on minisuperspace. 
  
  In the present cosmological model the minisuperspace is a $3D$ 
  space  $\{a,\phi,\psi\}$ (or $\{P,M,R\}$) and the associated conjugate 
  momenta to the variables are   
  \begin{eqnarray}
  p_{_{_P}}&=&\frac{\partial 
  	L}{\partial\dot{P}}=e^M\left(2A\dot{P}+C\dot{M}\right)=\mbox{Conserved ($\Sigma_P,~$ 
  	say)} \nonumber \\
   p_{_{_M}}&=&\frac{\partial L}{\partial \dot{Q}}=e^M\left(C\dot{P}+2B\dot{M}\right) 
  \nonumber\\  p_{_{_R}}&=&\frac{\partial L}{\partial 
  	\dot{R}}=-2D\dot{R}e^M=\mbox{Conserved ($\Sigma_R,~$ say).}
  \end{eqnarray}
  
  So, the Hamiltonian of the system ( also known as Hamiltonian's constraint) 
  takes the form,
    \begin{equation}
  H=e^{-M}\left[A_1 p_{_{_P}}^2 + A_2 p_{_{_M}}^2 + A_3 p_{_{_P}} p_{_{_M}} - A_4 p_{_{_R}}^2+V_0e^{2M}\right]
  \end{equation}
  where $A_1,~A_2,~A_3,~A_4~$ are constant and they are connected with $A,~B,~C$.
  Then the WD equation (which is the operator version of the above Hamiltonian 
  (51)) has the explicit form 
  \begin{equation}
  \left[-A_1e^{-M}\frac{\partial^2}{\partial P^2} - A_2e^{-M} 
  \frac{\partial^2}{\partial M^2} - A_3e^{-M}\frac{\partial^2}{\partial 
  	P\partial M}+A_4e^{-M}\frac{\partial^2}{\partial R^2}+V_0e^M\right] 
  \Psi(P,M,R)=0
  \end{equation}
  where $\Psi$ is termed as wave function of the universe. In the context of WKB 
  approximation the above wave function can be written as $\Psi(q^{\alpha})e^{i 
  	\delta_{\alpha}(q^{\alpha})}$ and consequently the WD equation becomes first 
  order non-linear partial differential equations which is the usual (null) 
  Hamilton-Jacobi(H-J) equation in the minisuperspace.
  
  Further, the general solution of WD equation can be obtained as the 
  superposition of eigen function of the above WD operator as\cite{rh} 
  \begin{equation}
  \Psi(P,M,R)=\int W(Q)\Psi(P,M,R,Q)dQ
  \end{equation} 
  here $\Psi$ is an eigen function of the WD operator and $W(Q)$ is weight 
  function of the conserved charge $Q$. Usually, it is preferable that 
  the wave function in quantum cosmology should be consistent with the classical 
  theory i.e., there should be a coherent wave packet with the good asymptotic 
  behavior in the minisuperspace and have a maximum around classical trajectory.
  
  In order to solve the above WD equation, we write the wave function in the form
  \begin{equation}
  \Psi(P,M,R)=\psi_1(P)\psi_2(M)\psi_3(R)\nonumber\\
  \end{equation}
  as P,R are the cyclic variables so operator version of the associated conserved charge take the form.
  \begin{eqnarray}
  -i\frac{\partial\psi_1(P)}{\partial P}&=&\Sigma_P\psi_1(P)\\
  -i\frac{\partial\psi_3(R)}{\partial R}&=&\Sigma_R\psi_1(R)
  \end{eqnarray}
  and the solution of the above equations (56),(57) are given below
  \begin{eqnarray}
  \psi_1(P)=\exp(i\Sigma_PP)\\
  \psi_3(R)=\exp(i\Sigma_RR)
  \end{eqnarray}
  then by substituting the value of (58) and (59) in the WD equation (55), explicit form of the wave function becomes
  \begin{eqnarray}
  \implies	\Psi=e^{i(\Sigma_PP+\Sigma_RR)}
  \left\{M_1e^{-\frac{B^*M}{A^*}}J_1\left[\frac{\sqrt{{(B^*)}^2-4A^*C^*}}{2A^*}, 
  \sqrt{\frac{V_0}{A^*}}e^M\right]\right.\nonumber\\\left.+M_2e^{-\frac{B^*M}{2A^*}}
  J_2\left[\frac{\sqrt{(B^*)^2-4A^*C^*}}{A^*}, \sqrt{\frac{V_0}{A^*}}e^M\right]\right\}
  \end{eqnarray}
  where $~A^*,B^*,C^*,D^*,E^*~$ are given below as
  \begin{eqnarray}
  A^*&=&\big(4AB^2C^2+16A^2B^3-8AB^2C^2 \big)\nonumber\\
  B^*&=&\big(AC^4+B(4AB-C+2AC)^2-C^3(4AB-C+2AC)\big)\nonumber\\
  C^*&=&-\frac{C^2(4AB-C)^2}{4D}\nonumber\\
  D^*&=&\big(-4BC^3A-(8AB^2+c^3)(4AB-C+2AC)-8AB^2C^2\big)\nonumber\\
  E^*&=&V_0C^2(4AB-C)^2
  \end{eqnarray}
  and $J_1~,~J_2$ are Bessel functions of 1st and 2nd kind.

\begin{figure}[h]
	\centering \includegraphics[height=5cm,width=7cm]{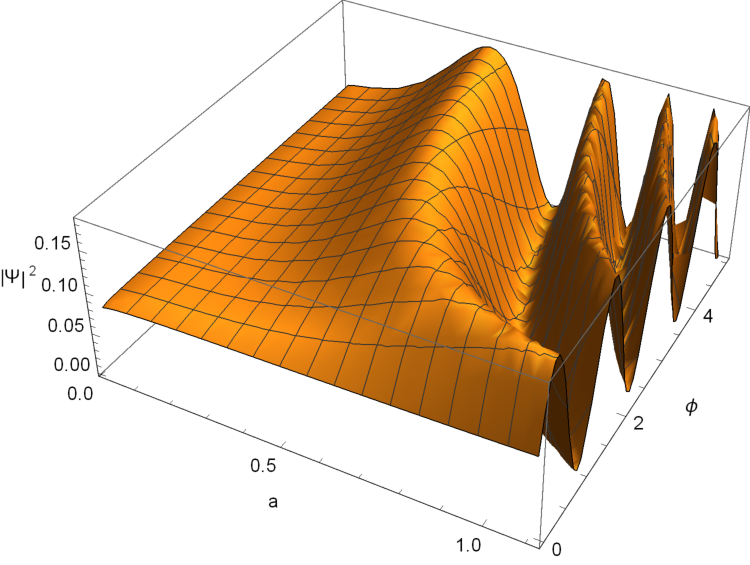}
	\caption{3D graphical presentation of $|\Psi|^2$ versus the scale factor $a$ and $\phi$ .}\label{f2}
\end{figure}

  Usually a consistent quantum cosmological model should predict the classical 
  solution at late time, while at early time the quantum model should be distinct 
  from the classical solution (i.e., singularity) to have a well defined era of 
  evolution. For the present problem we have plotted $|\Psi|^2$ against $a, \phi$ in FIG. \ref{f2}. From the figure it is clear that there is finite non-zero probability at zero volume . Hence the initial classical singularity (i.e., big-bang singularity) can not be remove by quantum description. 
  
  \section{Brief Discussion and Conclusion}
  The present work is an example where the Noether symmetry analysis has been 
  extensively used in a very complicated cosmological model namely coupled 
  multiscalar field model. In the context of classical cosmology, the symmetry 
  analysis not only determines the coupling function (instead of phenomenological
  choice) but also gives the classical solution by choosing the cyclic variable 
  in the augmented space. The graphical representation of the cosmological 
  parameters show that the evolution of the present model agrees qualitatively 
  with the observational predictions. The geometric symmetry of the physical 
  space shows that there is a homothetic vector field and there is no Noether 
  symmetry along the homothetic vector field. Further, the physical space is 
  conformally flat and the model is equivalent to a quintom model. In quantum 
  cosmology, the WD equation is highly coupled non-linear second order 
  (hyperbolic) partial differential equation for this complex cosmological model.
  The operator version of the conserved charge equations identify the oscillatory 
  part of the wave function and as a result the WD equation becomes solvable. 
  The graphical representation of the wave function has been shown in FIG. \ref{f2}. for various choices of the parameters involve and from the figure we see that the wave function has a finite value at $a=0$ i.e., there is a finite probability for the system to have zero volume. So the wave function does not vanish at the big bang singularity. Therefore, the quantum version 
  of the present cosmological model {can not remove the big-bang singularity  rather the initial classical singularities continue to be present at the quantum level. Finally, for this strange behavior an alternative interpretation of quantum mechanics: the de Bohm--Broglie interpretation of quantum mechanics may be useful} (\cite{r46} \cite{r47}).

  \section*{Acknowledgment}
  Author D.L. thanks UGC, Government of India for awarding Junior research Fellowship (JRF)  (ID: 211610000075).


\frenchspacing
\begin{thebibliography}{58}
 \bibitem{r1} C. Brans and R.H. Dicke,  {\it Phys.
Rev.} {\bf 124}, 195 (1961).

\bibitem{r2} V. Faraoni, Cosmology in Scalar-Tensor Gravity, Fundamental Theories of Physics {\bf vol. 139},
Kluwer Academic Press: Netherlands, (2004).

\bibitem{r3} M. Hohmann, {\it Phys. Rev. D} {\bf 98}, 064002 (2018).

\bibitem{r4} S.V. Chernov, {\it Quantum Matters} {\bf 2}, 71 (2013).

\bibitem{r5} A.M. Perelomov, {\it Phys. Rept.} {\bf 146}, 135 (1987).

\bibitem{r6} S.V. Chernov, {\it Gravit. Cosmol.} 1, {\bf 91} (1995).

\bibitem{r7} P. Christodoulidis, D. Roest and E.I. Sfakianakis, 
{\it JCAP} {\bf 11}, 002 (2019).

\bibitem{r8}  P. Christodoulidis, D. Roest and E.I. Sfakianakis,  {\it JCAP} {\bf 12}, 059 (2019).

\bibitem{r9}  C.-B. Chen and J. Soda, {\it JCAP} {\bf 09}, 026 (2021).

\bibitem{r10}
A. Paliathanasis and G. Leon, {\it Class. Quantum Grav.} {\bf 38}, 075013 (2021).


\bibitem{r11} A. Paliathanasis and G. Leon, {\it Eur.Phys. J. Plus} {\bf 137}, 165 (2022).


\bibitem{r23} M. Tsamparlis and A. Paliathanasis, {\it J. Phys. A} {\bf 44}, 175202 (2011).


\bibitem{r24} P. G. L. Leach, {\it Australian Mathematical Society Lecture Series} {\bf 22}, 12 (2009).


\bibitem{r25} G. Bluman and S. Kumei, {\it Symmetries and Differential Equations} {\bf (Springer-Verlag, N. Y. 1989)};

\bibitem{r44} H. Stephani, {\it Differential Equations: Their Solutions Using Symmetry} {\bf (Camb. Univ.Press, Cambridge, England,1989)};

\bibitem{r45} P. J. Olver, {\it Applications Of Lie Groups to Differential Equations} {\bf (Springer, N. Y. 1986)}.


\bibitem{r26} A. V. Aminova, {\it Mat. Sb.} {\bf 186}, 1711 (1995).


\bibitem{r27} A. V. Aminova and N. A. M. Aminov, {\it Tensor Newser.} {\bf 62}, 65 (2000).


\bibitem{r28} T. Feroze, F. M. Mahomed and A. Qadir, {\it Nonlinear Dynamics} {\bf 45}, 65 (2006).


\bibitem{r29} M. Tsamparlis and A. Paliathanasis, {\it Gen.Relt.Grav.} {\bf 42}, 2957 (2010).


\bibitem{r36} S. Dutta, M. Lakshmanan and S. Chakraborty,  {\it Annals of Phys.} {\bf 407}, 1 (2019).	


\bibitem{r37} S. Dutta, M. Lakshmanan and S. Chakraborty,  {\it Phys. of Dark. Univ.} {\bf 32}, 100795 (2021).

\bibitem{r35} M. Tsamparlis and A. Paliathanasis, {\it Class. Quant.Grav.} {\bf 29}, 015006 (2012).

\bibitem{r16} S. Dutta and S. Chakraborty,  {\it Int. J. Mod. Phys. D} {\bf 25}, 1650051 (2016).


\bibitem{r17} S. Dutta, M. M. Panja and S. Chakraborty,  {\it Gen. Relt. Grav.} {\bf 48}, 54 (2016).

\bibitem{r44a} S. Capozziello, A. Stabile and A. Troisi, {\it Class. Quant. Grav.} {\bf 24}, 2153 (2007).

\bibitem{rh} Gravitation in astrophysics, Cargèse, 1986 / edited by B. Carter and J.B. Hartle.(New York : Plenum Press, 1986.)p-1-399.

\bibitem{rA} A. Paliathanasis, {\it Universe} {\bf 8}, 325 (2022).
\bibitem{Bhaumik:2022adi} R.~Bhaumik, S.~Dutta, and S.~Chakraborty,  {\it Int. J. Mod. Phys. A} {\bf 37 }, 2250018 (2022).
\bibitem{r42} N. Banerjee, S. Das, K. Ganguly, {\it Pramana} {\bf 74},481 (2010).

\bibitem{r43} H. Sheikhahmadi et.al.,{\it Eur. Phys. J. C} {\bf 79}, 1038 (2019).

\bibitem{r46}  S. Chakraborty, {\it Int. J. Mod. Phys. D} {\bf 10},943 (2001).

\bibitem{r47} Jr. Roberto Colistere an J.C. Fabris, {\it Phys. Rev. D} {\bf 57}, 4707 (1998).



  
    
   
  


\end{thebibliography}
\end{document}